\documentstyle[11pt]{article} 

\def\ds{\displaystyle}
\begin{document} 
\centerline{\bf Statistical Physics on the space ($x,v$) for dissipative systems and study of an ensemble 
}
\centerline{\bf of harmonic oscillators in a weak linear dissipative medium}
\vskip2pc
\centerline{G.V. L\'opez$^1$, P. L\'opez$^1$,  and X. E. L\'opez$^2$}
\vskip0.5pc
\centerline{$^1$Departamento de F\'{\i}sica, Universidad de Guadalajara}
\centerline{Apartado Postal  4-137, 44410 Guadalajara, Jalisco, M\'exico}
\vskip0.5pc
\centerline{$^2$Facultad de Ciencias, Universidad Nacional Aut\'onoma de M\'exico}
\centerline{Apartado Postal 70-348, Coyoac\'an 04511, M\'exico D.F.} 
\vskip2pc
\centerline{PACS: 05.20.Gg, 05.30.Ch, 05.20.-y, 05.30.-d}
\vskip2cm
\centerline{ABSTRACT}
\vskip1pc\noindent
We use the phase space position-velocity ($x,v$) to deal with the statistical properties of
velocity dependent dynamical systems, like dissipative ones. Within this approach, we study the
statistical properties of an ensemble of harmonic oscillators in a linear weak dissipative media. Using
the Debye model of a crystal, we calculate at first order in the dissipative parameter the entropy, free
energy, internal energy, equation of state and specific heat using the classical and quantum approaches.
for the classical approach we found that the entropy, the equation of state, and the free energy depend
on the dissipative parameter, but the internal energy and specific heat do not depend of it. For the
quantum case, we found that all the thermodynamical quantities depend on this parameter.
\vfil\eject\noindent 
{\bf 1. Introduction}
\vskip0.5pc\noindent
Our foundation of classical and quantum statistical mechanics [1] is based on  the
Hamiltonian formalism [3] of conservative systems. These systems are particular cases of much more
general ones called autonomous systems which are those where the total force acting on the particle
does not depend explicitly on time, otherwise they are called non-autonomous. For conservative systems,
the Hamiltonian is a constant of motion of the system, and, in
principle, one can  calculate all the thermodynamic characteristics  of an ensemble of N particles
governed by this Hamiltonian,  through the acknowledge of its associated partition function. For
dynamical systems which depend explicitly on the velocity, for example dissipative systems. There are
mainly two approaches to study these systems. The first one uses the kinetic equation (Vlasov or Boltzmann
equations) to find the density distribution  and, then, to calculate the desired statistical property. The
other one tries to find an effective or phenomenological Hamiltonian and, then, to estimate the desired
thermodynamic quantities from the stationary statistical mechanics (statistical ensemble)[5]. This last
approach is possible since for  autonomous systems sometimes is possible to find time-independent
constant of motion, Lagrangian, and Hamiltonian. However, it is known that the Lagrangian and Hamiltonian
formulation have some problems. First, the Lagrangian (therefore the Hamiltonian) may not exist for some
dynamical systems [6]. Second, There could be two completely different Hamiltonians which describe the
same classical mechanical system, but completely different quantum and statistical systems [7]. Third,
time-explicitly dependent systems have an ambiguous Lagrangian and Hamiltonian formulation [8]. Fourth,
Ambiguity Lagrangian and Hamiltonian formulation happens even for the harmonic oscillator [9] (where a
problem of consistency of units arises). Finally, there are autonomous systems where one can not have
explicitly the Hamiltonian of the system because the inverse relation $v=v(x,p)$ can not be given from
the original definition of $p=p(x,v)$ [10]. This last problem happens particularly on dissipative systems
and is the main reason one would like to have an alternative approach in mechanics [11], statistical and
quantum mechanics to deal with them. 
\vskip0.5pc\noindent
In this paper, we take the original ideas of Maxwell in statistical mechanics and Heisenberg in quantum
mechanics  to study these areas from the point of view of coordinates and velocities rather than
coordinates and generalized linear momentum. We apply this approach to the  ensemble of linear
oscillators within a linear dissipative media, and we will do this from the phenomenological point of
view and at first order in the dissipation parameter in the constant of motion. 
\vskip2pc\noindent
\leftline{\bf 2. Dynamical equation in the space ($x,v$)}
\vskip1pc\noindent
In this section, we restrict ourselves to one-dimensional single particle motion to show the main ideas.
The approach will be extended immediately to N independent particles moving in a three-dimensional
space. The equations of motion of a particle moving under a time-independent force can be written as the
following autonomous dynamical system
$$\dot x=v\ ,\eqno(1a)$$
and
$$\dot v={f(x,v)\over m}\ ,\eqno(1b)$$
where $m$ is the mass of the particle, $x$ and $v$ represent its position and velocity, and $f(x,v)$ is
the total force acting on the particle.  A constant of motion of this system is a function $K(x,v)$ such
that $dK/dt=0$, that is, it satisfies the following equation
$$v{\partial K\over\partial x}+{f(x,v)\over m}{\partial K\over\partial v}=0\ .\eqno(2)$$
Given this constant of motion, the Lagrangian of the system is given by [15]
$$L=v\int{K(x,v)\over v^2}dv+A(x)v\ ,\eqno(3)$$
where $A(x)$ is an arbitrary function, and the term $A(x)v$ represents the gauge  of the Lagrangian.
Thus, the generalized linear momentum is given by
$$p={\partial L\over\partial v}\ ,\eqno(4)$$
and it is here where our analysis starts. Suppose now that it is not possible to get $v=v(x,p)$ from
(4). Then, the Hamiltonian of the system, $H=vp-L(x,v)$, can not be given explicitly, and it will be
given in implicit form through the constant of motion. Therefore, this constant of motion (with units of
energy)  can help us to avoid this an other problems already mentioned in the introduction.
\vskip0.5pc\noindent
Consider $N$ particles moving in the three-dimensional space and under  a time-independent force acting on
them (autonomous system). The motion of these particles is restricted on the hypersurface of the
space $\Re^{6N}$ defined by the time-independent constant of motion $K({\bf x},{\bf v})$, where ${\bf
x},{\bf v}\in\Re^{3N}$. Thus, for a canonical ensemble of $N$-particles , the classical partition
function would be given by
$$Z={m^{3N}\over h^{3N}\eta}\int e^{-\beta K({\bf x},{\bf v})}d{\bf x}d{\bf v}\ ,\eqno(5)$$
where $m$ is the mass of the particles, $h$ is the Plank's constant, $\eta=1$ for distinguishable particles
and $\eta=N!$ for non distinguishable particles,  $\beta=1/kT$ with $k$ being the Boltzmann's constant and
$T$ the temperature of the system, and $d{\bf x} d{\bf v}$ is the measure in the space $\Re^{6N}$,
$$d{\bf x} d{\bf v}=\prod_{j=1}^{3N}dx_j dv_j\ .\eqno(6)$$
For a quantum canonical ensemble of $N$ particles in a batch temperature $T$, the quantum partition
function would be
$$Z=\sum_i\omega_ie^{-\beta E_i}\ ,\eqno(7)$$
where $\omega_i$ represents the degeneration of the eigenvalue $E_i$, where this eigenvalues comes from
the solution of the equation
$$\widehat{K}(\widehat{\bf x},\widehat{\bf v}) \psi_i=E_i\psi_i\ .\eqno(8a)$$
$\widehat{K}$, $\widehat{\bf x}$ and $\widehat{\bf v}$ are the Hermitian operator associated to the
constant of motion $K({\bf x},{\bf v})$, the position ${\bf x}$, and the velocity ${\bf v}$. $\psi_i({\bf
x})$ is the eigenfunction which is related with the wave function $\Psi({\bf x},t)$ of the Schr\"odinger
equation,
$$i\hbar{\partial\Psi({\bf x},t)\over\partial t}=\widehat{K}(\widehat{\bf x},\widehat{\bf v}) 
\Psi_i({\bf x},t)\ ,\eqno(8b)$$
as
$$\Psi({\bf x},t)=\sum_iC_ie^{-iE_it/\hbar}\psi_i({\bf x})\ ,\eqno(8c)$$
where $|C_i|^2$ represents the probability that the system be on the state $\psi_i$. The position and
velocity operators are defined and related by the following expressions
$$\widehat{x_j}=x_j\ ,\hskip2pc \widehat{v_j}=-{i\hbar\over m}{\partial\over\partial x_j}\ ,\hskip2pc
[x_l,\widehat{v_j}]={i\hbar\over m}\delta_{lj},\hskip2pc [x_i,x_j]=[v_i,v_j]=0\ .\eqno(8d)$$
In this way, with Eq. (5) and Eq. (7) it is possible, in principle, to study the thermodynamic
characteristics of an ensemble of $N$ particles of any autonomous system through the classical or quantum
canonical partition functions. Of course, this approach is reduced to that one of the Hamiltonian
formalism  whenever this one exists explicitly [11]. Note that the same approach can be made for any
other type of ensemble.
\vskip3pc\noindent
\leftline{\bf 3. Constant of motion}
\vskip0.5pc\noindent
In reference [12], a constant of motion $K_{\alpha}(x,v)$ was given for the dynamical system
$$\dot x=v\eqno(9a)$$
and
$$\dot v=-\omega^2x-{\alpha\over m}v\ ,\eqno(9b) $$
where $\omega$ is the free natural frequency of oscillations, $m$ is the mass of the particle,  and
$\alpha$ is the coefficient of the dissipative force which arises phenomenologically as an average
effect of the interaction with the particles of the medium. The constant of motion of this system is
given by
$$K_{\alpha}(x,v)={m\over 2}\biggl(v^2+2\omega_{\alpha}x v+\omega^2x^2\biggr)
e^{-2\omega_{\alpha}G(v/x,\omega,\omega_{\alpha})}\ ,\eqno(10a)$$
where $\omega_{\alpha}$ and the function $G$ are defined as
$$\omega_{\alpha}={\alpha\over 2m}\ ,\eqno(10b)$$
and
$$G(\xi,\omega,\omega_{\alpha})=
\cases{{1\over\ds 2 \sqrt{\omega_{\alpha}^2-\omega^2}}
\ln\Biggl[{\ds\omega_{\alpha}+\xi-\sqrt{\omega_{\alpha}^2-\omega^2}\over\ds
\omega_{\alpha}^2+\xi+\sqrt{\omega_{\alpha}^2-\omega^2}}\Biggr]& if $\omega^2<\omega_{\alpha}^2$\cr\cr
{ 1\over\ds \omega_{\alpha}+\xi}& if $\omega^2=\omega_{\alpha}^2$\cr\cr
{1\over\ds\sqrt{\omega^2-\omega_{\alpha}^2}}
\arctan\Biggl({\ds\omega_{\alpha}+\xi\over\ds\sqrt{\omega^2-\omega_{\alpha}^2}}\Biggr)& if
$\omega^2>\omega_{\alpha}^2$\cr}\eqno(10c)$$
This constant of motion has the following limit
$$\lim_{\alpha to 0}K_{\alpha}(x,v)=K_0(x,v)={1\over 2}mv^2+{1\over 2}m\omega^2 x^2\ .\eqno(11)$$
For very weak dissipation ($\omega_{\alpha}\ll \omega$) and at first order in $\omega_{\alpha}$, the
constant of motion can be written as
$$K(x,v)=K_0(x,v)+{\omega_{\alpha}\over\omega}\biggl[m\omega x v-2K_0(x,v)
\arctan\biggl({v\over\omega x}\biggr)\biggr]\ .\eqno(12)$$
In principle, one can get the Lagrangian of the system through the expression (3), and from this
Lagrangian one could get the generalized linear momentum of the system $p=p(x,v)$. However, it is not
possible to know
$v=v(x,p)$, even  for the very weak dissipation case. Thus, the Hamiltonian is given implicitly through
the above constant of motion.
\vskip0.5pc\noindent
The quantization of (12) was carried out in reference [13], where the modification of the eigenvalues of
the harmonic oscillator ($E_n^{(0)}=\hbar\omega(n+1/2)$) were given at first order in perturbation
theory as
\begin{eqnarray*}
E_n(\omega)&=&\hbar\omega\left(n+{1\over 2}\right)\biggl(1-{\pi\omega_{\alpha}\over\omega}\biggr)\\
& &-{\hbar\omega_{\alpha}^2\over\omega}\left[({2\over 3}n
+{1\over 4})-\sum_{l=0}^{\infty}\sum_{s=0}^{2l+1}
\pmatrix{-1/2\cr l}^2\pmatrix{2l+1\cr s}^2{(2n-2l-s)^2\over (2l+1)^2(2l+1+s)}\right]\ ,
\end{eqnarray*}
$$\eqno(13)$$
where the correction has been made up to second order in $\omega_{\alpha}$. 
\vskip1pc\noindent
\leftline{\bf 4. Classical approach}
\vskip0.5pc\noindent
Consider an ensemble of $N$ particles moving independently which their equation of motion is given by the
following dynamical system
$${d\over dt}\pmatrix{x_j\cr v_j}=\pmatrix{0&1\cr-\omega_j^2&-{\alpha\over m}}\pmatrix{x_j\cr v_j}\ ,
\hskip3pc j=1,\dots,3N\eqno(14)$$
where $\omega_j$ is the natural free frequency of oscillations. The dissipative coefficient is the same
of all oscillators since they are inside the same medium. One can think, for example, in a
crystal which is inside a bosonic or fermionic medium where the wave length of the particles of this medium 
is less than the separation of the crystal
components [16]. The average interaction of the crystal components with
the particles of the medium may be modeled by Eq. (14). In Astrophysics, one could think about the core of
a neutron star as a superconducting lattice of protons inside a dissipative medium generated by the huge
amount of neutrinos (antineutrinos) appearing from the weak decay of neutrons and protons. Since the
density in the core of these stars is very big ($10^{15}gr/cm^3$)[17] the interaction with neutrinos
(antineutrinos) is not  so neglectful. Thus, maybe the local motion of an ensemble of  nucleons in the
core could be model by Eq. (14). 
\vskip0.5pc\noindent
Let us see first the implications of using classical canonical partition function to study the
thermodynamical characteristics of the system. To do this, let us make the following change of variable
$$x_j=\sqrt{2J_j\over m\omega_j}\cos{\phi_j}\ ,\hskip3pc\hbox{and}\hskip3pc 
vi=\sqrt{2\omega_jJ_j\over m}\sin(\phi_j)\ ,\eqno(15)$$
where the variables $\phi_j$ and $J_j$ have the following variation $\phi_j\in[0,2\pi]$ and
$J_j\in[0,\infty)$. Then, from (12), the jth-constant of motion and measure are given with this
coordinates as
$$K_j=\omega_jJ_j-\omega_{\alpha}J_j(\sin{2\phi_j}-2\phi_j)\ ,\eqno(15a)$$
and 
$${dx_jdv_j}={1\over m}dJ_jd\phi_j\ .\eqno(15b)$$
In this way, the partition function (5) is written as ($\eta=1$ since the particles in the crystal are
distinguishable)
$$Z={m^{3N}\over h^{3N}}\prod_{j=1}^{3N}\int e^{-\beta K_j(x_j,v_j)}dx_jdv_j=\prod_{j=1}^{3N}
{1\over h}\int e^{-\beta J_j[\omega_j-\omega_{\alpha}(\sin{2\phi_j}-2\phi_j]}dJ_jd\phi_j
\eqno(16)$$ 
which can be integrated, bringing about the following result
$$Z=\prod_{j=1}^{3N}{I\left({\omega_{\alpha}/ \omega_j}\right)\over\beta h\omega_j}\ ,\eqno(17a)$$
where the function $I(\xi)$  has been defined as
$$I(\xi)=\int_0^{2\pi}{d\phi\over{1-\xi(\sin{2\phi}-2\phi)}}\ .\eqno(17b)$$
Fig. 1 shows the behavior of this function as a function of $\omega_{\alpha}/\omega$. Note that
$I(0)=2\pi$, and  that $\omega_{\alpha}/\omega$ must be less than the unit, according to our
approximations of expression (5).  From (17a), one gets
$$\ln Z=-3N\ln{h}+\sum_{j=1}^{3N}\ln\left({I(\omega_{\alpha}/\omega_j)\over\beta\omega_{j}}\right)\
.\eqno(18)$$
Since $N$ is big and there are $3N$ normal modes of oscillations in the crystal, one can assume
continuity in the frequency spectrum  and write (18)  as
$$\ln Z=U_0+\int_0^{\infty} \ln\left({I(\omega_{\alpha}/\omega)\over\beta\omega}\right) g(\omega)~d\omega
\ ,\eqno(19a)$$
where $U_0$ has been defined as $U_0=-3N\ln{h}$, and $g(\omega)$ is the density spectral which must
satisfied the condition
$$\int_0^{\infty}g(\omega)~d\omega=3N\ .\eqno(19b)$$
Using the Debye's model of solids [14], the spectral density is given by
$$g(\omega)=\cases{{\ds 9N\omega^2\over\ds\omega_{D}^3}& if $0\le\omega\le\omega_D$\cr\cr
0& if $\omega>\omega_D$\cr}\ ,\eqno(20a)$$
where $\omega_D$ is the Debye's frequency of the solid which is defined by the cutoff frequency such that
$$\int_0^{\omega_D}g(\omega)~d\omega=3N\ .\eqno(20b)$$
and it is given by
$$\omega_D=\left({3N\over 4\pi V}\right)^{1/3}v_c\ .\eqno(20c)$$
The variable $V$ represents the volume of the solid, and $v_c$ is the average velocity of the elastic
waves in the solid which is given in terms of the longitudinal ($v_l$) and transversal ($v_t$) waves as
$${3\over v_c}={2\over v_t^3}+{1\over v_l^3}\ .\eqno(20d)$$
Substituting (20a) in (19a), it follows that
$$\ln Z=U_0+{9N\over\omega_D^3}\int_0^{\omega_D}\omega^2
\ln\left({I(\omega_{\alpha}/\omega)\over\beta\omega}\right)~d\omega
\ .\eqno(21)$$
In this way, one can  calculate the thermodynamic characteristics of the system. The internal
energy of the system, its entropy, its specific heat, its equation of state, and its free energy are given
by
$$U=-{\partial\ln Z\over\partial\beta}=3NkT\ ,\eqno(22a)$$
$$S=k\biggl[U_0+{9N\over\omega_D^3}\int_0^{\omega_D}\omega^2
\ln\left({I(\omega_{\alpha}/\omega)\over\beta\omega}\right)~d\omega+3N\biggr]\ ,\eqno(22b)$$
$$C_V=\left({\partial U\over\partial T}\right)_V=3Nk\ ,\eqno(22c)$$
$$p={3NkT\over V}\ln\left({I(\omega_{\alpha}/\omega_D)\over\beta\omega_D}\right)-
{9N\over V\omega_D^3}\int_0^{\omega_D}\omega^2
\ln\left({I(\omega_{\alpha}/\omega)\over\beta\omega}\right)~d\omega\ ,\eqno(22d)$$
and
$$F=-kT\left[U_0+{9N\over\omega_D^3}\int_0^{\omega_D}\omega^2
\ln\left({I(\omega_{\alpha}/\omega)\over\beta\omega}\right)~d\omega\right]\ .\eqno(22c)$$
As one can see, the internal energy and the specific heat of the ensemble of oscillators do not depend on
the dissipative media, at first order in the dissipation parameter. We must remain here that the whole
system is made up of the crystal and the medium. So, in the above quantities one needs to add the
contribution coming purely from the medium. On the other hand, we have used the classical canonical
statistical partition function and the Debye's model for our study of the thermodynamic quantities of the
system. However, this is somewhat a little bite incorrect since Debye's model works fine to relative low
temperatures, and classical canonical partition function is expected to work fine to relatively high
temperatures. Thus, let us make everything consistent by using quantum canonical partition function.
\vfil\eject
\vskip2pc
\leftline{\bf 5. Quantum approach}
\vskip0.5pc\noindent
In this case, using (7), (20a), the condition (20b), and the same hypothesis of continuity in the
frequencies, one has the following expression 
\begin{eqnarray*}
\ln Z&=&\ln\prod_{j=1}^{3N}\left(\sum_ne^{-\beta E_n(\omega_j)}\right)\\
&=&\sum_{j=1}^{3N}\ln\left(\sum_ne^{-\beta E_n(\omega_j)}\right)\approx
\int_0^{\infty}\ln\left(\sum_ne^{-\beta E_n(\omega)}\right)g(\omega)~d\omega\\
&=&{9N\over\omega_D^3}\int_0^{\infty}\omega^2\ln\left(\sum_ne^{-\beta E_n(\omega)}\right)~d\omega\ ,
\end{eqnarray*}
$$\eqno(23)$$
Now, using (13) at first order in the dissipation parameter, one has
$$\sum_ne^{-\beta E_n(\omega)}={\lambda_{\alpha}e^{-\beta\hbar\omega/2}\over 1-\lambda_{\alpha}^2
e^{-\beta\hbar\omega}}\ ,\eqno(24a)$$
where $\lambda_{\alpha}$ has been defined as
$$\lambda_{\alpha}=e^{\beta\pi\hbar\omega_{\alpha}/2}\ .\eqno(24b)$$
Thus,  Eq. (23) is written in the following way
$$\ln Z={9N\over\omega_D^3}\int_0^{\omega_D}\omega^2\ln\left({\lambda_{\alpha}e^{-\beta\hbar\omega/2}\over 1-\lambda_{\alpha}^2
e^{-\beta\hbar\omega}}\right)~d\omega\ .\eqno(25)$$
Therefore, the internal energy, the entropy, the specific heat, the equation of state and the free energy
of the system are given by
$$U={9N\over 2\omega_D^3}\int_0^{\omega_D}\omega^2(\hbar\omega-\hbar\omega_{\alpha}\pi)
\coth\left({\beta\hbar\omega\over 2}-{\beta\hbar\omega_{\alpha}\pi\over 2}\right)~d\omega\ ,\eqno(26)$$

$$S={9Nk\over\omega_D^3}\int_0^{\infty}\omega^2\left[\ln\left({\lambda_{\alpha}e^{-\beta\hbar\omega/2}\over
1-\lambda_{\alpha}^2e^{-\beta\hbar\omega}}\right)+
{\beta\over 2}(\hbar\omega-\hbar\omega_{\alpha}\pi)
\coth\left({\beta\hbar\omega\over 2}-{\beta\hbar\omega_{\alpha}\pi\over 2}\right)\right]d\omega\
,\eqno(27)$$
$$C_V={9N\over 4kT^2\omega_D^3}\int_0^{\omega_D^2}{\omega^2(\hbar\omega-\hbar\omega_{\alpha}\pi)~d\omega
\over\sinh^2\left({\beta\hbar\omega\over 2}-{\beta\hbar\omega_{\alpha}\pi\over 2}\right)}\ .\eqno(28)$$

$$p={3NkT\over V}\ln\left({\lambda_{\alpha}e^{-\beta\hbar\omega_D/2}\over
1-\lambda_{\alpha}^2e^{-\beta\hbar\omega_D}}\right)-
{9NkT\over V\omega_D^3}\int_0^{\infty}\omega^2\ln\left({\lambda_{\alpha}e^{-\beta\hbar\omega/2}\over
1-\lambda_{\alpha}^2e^{-\beta\hbar\omega}}\right)~d\omega\ ,\eqno(29)$$
and
$$F=-{9NkT\over\omega_D^3}\int_0^{\infty}\omega^2\ln\left({\lambda_{\alpha}e^{-\beta\hbar\omega/2}\over
1-\lambda_{\alpha}^2e^{-\beta\hbar\omega}}\right)~d\omega\ .\eqno(30)$$
Fig. 2 shows the variation of $C_V/Nk$ as a function of the temperature for several values the
dissipative parameter $\omega_{\alpha}$. The effect at first order of the dissipation is to increases
the specific heat a low temperatures.
\vskip3pc
\leftline{\bf 5. Conclusions}
\vskip0.5pc\noindent
We have used  the phase space (${\bf x},{\bf v}$) to study the statistical properties of an ensemble of
harmonic oscillators within a dissipative medium, where its effect on the oscillators is to create a
linear velocity depending force. We have made the study at first order in the dissipation parameter for
classical partition function and quantum partition function, taking the Debye's model of solids as an
example for possible applications. The classical canonical partition function lead us to have an internal
energy and specific heat which are independent on the dissipation. However, quantum canonical partition
function brings about dependence on the dissipation for all thermodynamical variables of the
system. It is our feeling that the use of the space (${\bf x},{\bf v}$) for statistical physics studies
has less restrictions that the space (${\bf x},{\bf p}$).

\vskip5pc\noindent
\vfil\eject
\leftline{\bf Figure Captions}
\vskip1pc\noindent
Fig. 1 Function $I(\omega_{alpha}/\omega)$.
\vskip0.5pc\noindent
Fig. 2 Specific heat with $\theta_D=\hbar\omega_D/k=150~K$ (corresponding to Sodium). (1):
$150\omega_{\alpha}\pi/\omega_D=0.0$, (2): $150\omega_{\alpha}\pi/\omega_D=0.5$, (3):
$150\omega_{\alpha}\pi/\omega_D=1.0$, (4): $150\omega_{\alpha}\pi/\omega_D=2.0$, (5):
$150\omega_{\alpha}\pi/\omega_D=5.0$, (6): $150\omega_{\alpha}\pi/\omega_D=10.0$
\vskip0.5pc\noindent
\vfil\eject
\leftline{\bf References}
\obeylines{
1. R. Kubo, {\it Statistical Mechanics}, North-Holland, (1999).
2. A. Messiah, {\it Quantum Mechanics}, John Wiley and Sons, (1961) Vol.I,II. 
3. H. Goldstein, {\it Classical Mechanics}, Addison-Wesley, M.A., (1950)
4. G. L\'opez, IL Nuo. Cim. B, {\bf 115}, 2 (2000) 137. 
5. G. L\'opez, M. Murgu\'{\i}a and M. Sosa, Mod. Phys. Lett. B, {\bf 11},14 (1997) 625.
6. J. Douglas, Trans. Amer. MAth. Soc., {\bf 50} (1941) 71.
7. G. L\'opez, quant-ph/0504167 (2005).
8. G. L\'opez and J.I. Hern\'andez, Ann. Phys., {\bf 193}, 1(1989) 1.
9. G.L\'opez, Rev. Mex. Fis., {\bf 48} (2002) 10.
\quad G. L\'opez, Int. Jour. Theo. Phys., {\bf 37}, 5 (1998) 1617.
10. G. L\'opez, L.A. Barrera, Y. Garibo, H. Hern\'andez, J.C. Salazar and 
\quad C.A. Vargas, Int. Jour. Theo. Phys.,{\bf 43}, 10 (2004) 2009.
11. G. L\'opez, M. Murgu\'{\i}a and M. Sosa, Mod. Phys. Lett. B, {\bf 11} 14 (1997) 625.
12. G. L\'opez, Ann. Phys. {\bf 251}, 2 (1996) 372.
13. G.L\'opez and P. L\'opez, Int. Jour. of Theo. Phys.,{\bf 45}, 4 (2006) 734.
14. G.H. Wannier, {\it Statistical Physics}, Dover, (1966). Chap. 13.
\quad K. Huang, {\it Statistical Mechanics}, John Weley and Sons, (1987). Chap. 12.
\quad R.K. Pathrio,{\it Statistical Mechanics}, Butterworth Heinemann, (1996) Chap. 7.
\quad F. Schwabl, {\it Statistical Mechanics}, Springer-Verlag, (2002) Chap. 4.6
\quad D.A. Mc Quarrie, {\it Statistical Mechanics}, Harper-Collins, (1976) Chap. II.3
\quad M. Toda, R. Kubo and N. Sait\^o, {\it Statistical Physics},
\quad Springer-Verlag, (1998) vol. I, chap. 3.
15. J.A. Kobussen, Acta. Phys. Austr., {\bf 51} (1979) 193.
\quad C. Leuber, Phys. Lett. A, {\bf 86} (1981) 2.
\quad See also [8].
16. W. Jones and N.H. March, {\it Theoretical Solid State Physics}, Dover, (1985).
17. F. \"Ozel, Nature, {\bf 441}, 29 (2006) 115.
\quad L.Z. Fang and R. Ruffini, {\it Basic Concepts of Relativistic Astrophysics},
\quad World Scientific (1983). Chap. 3-4.
}

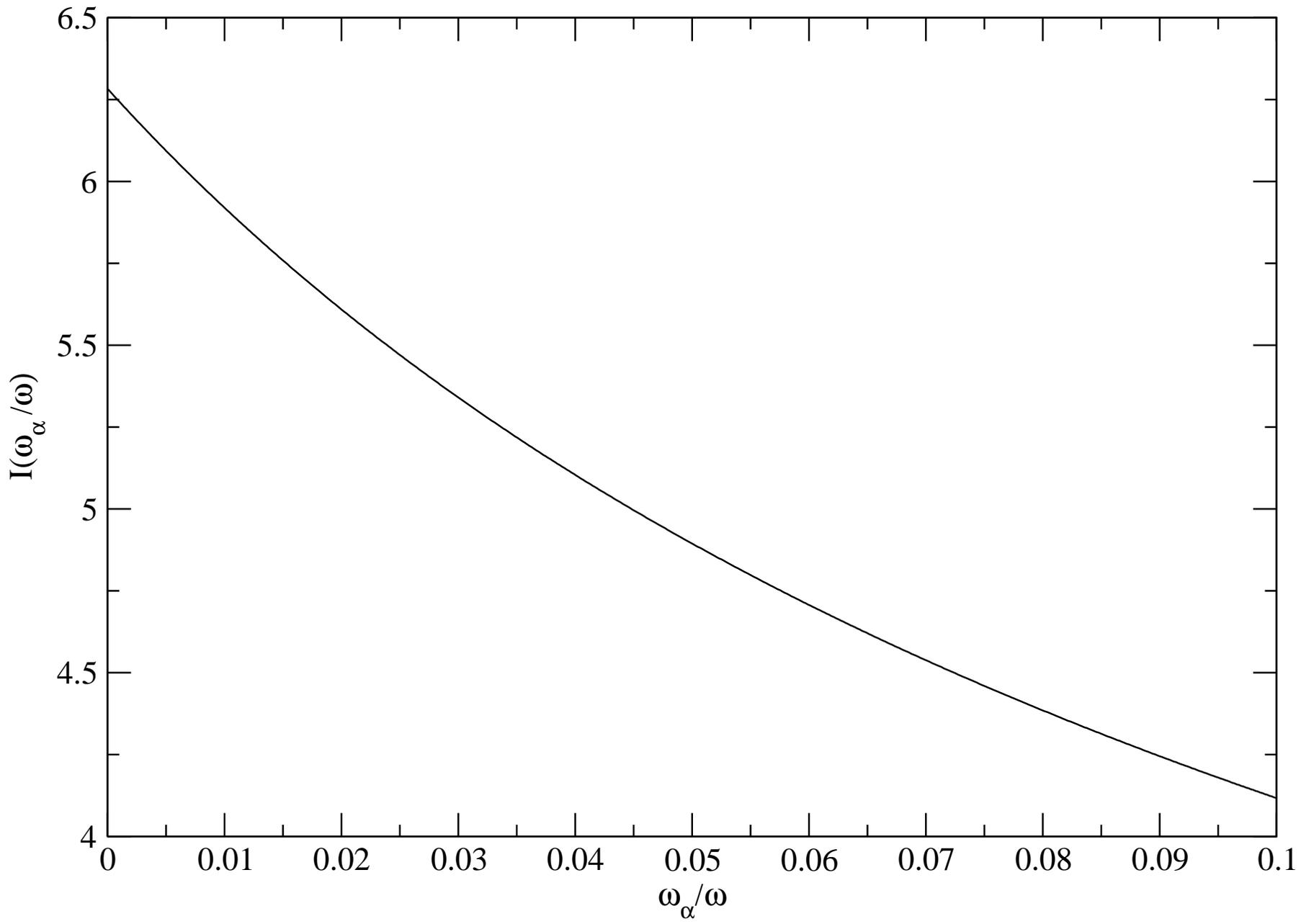
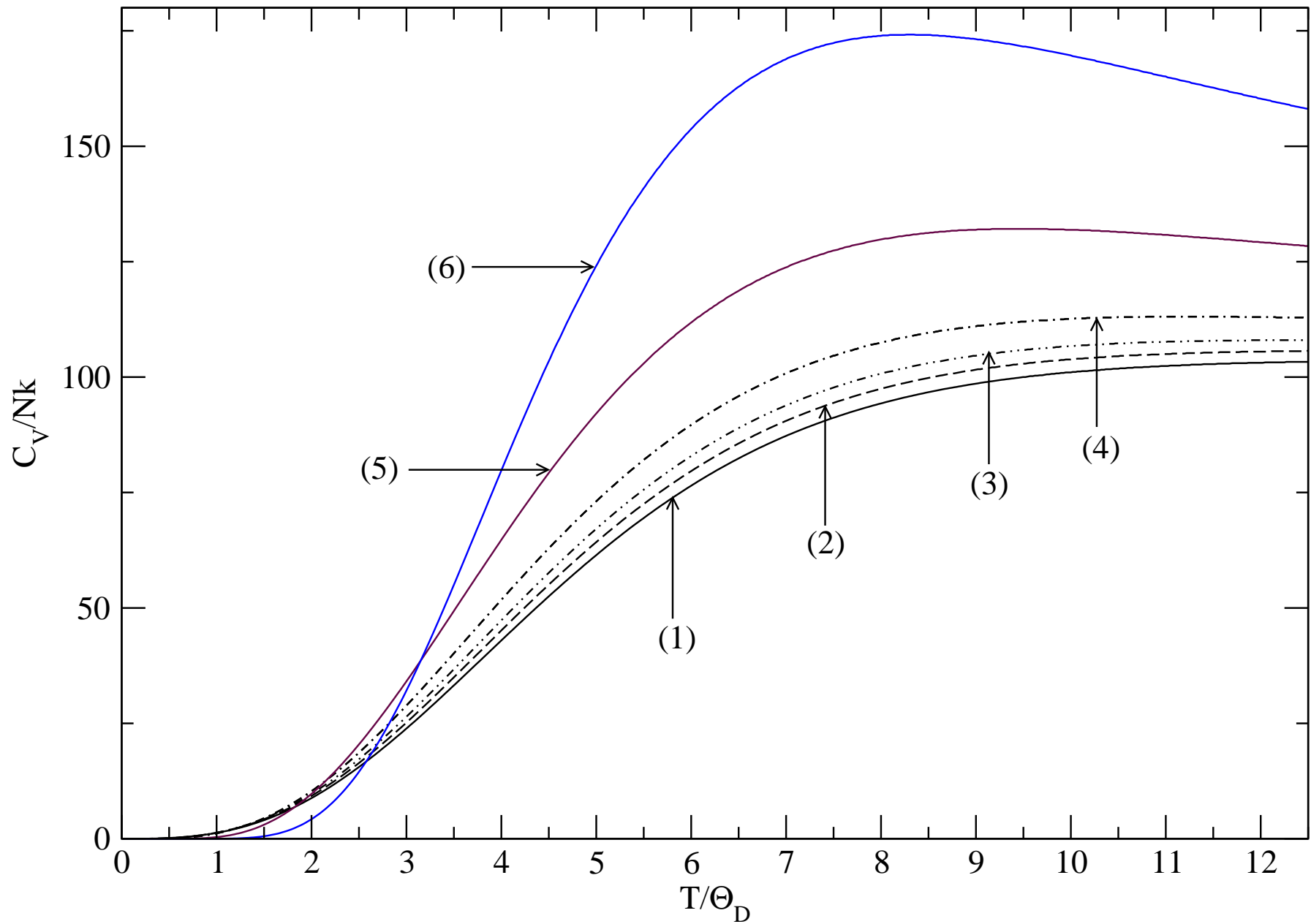
\end{document}